\documentclass[twocolumn]{revtex4-1}

\usepackage{graphicx,amsmath,amssymb,amsfonts,latexsym,color,dcolumn,bm,epsfig,subfigure}
\usepackage[plainpages=false,hyperfootnotes=false,colorlinks=false]{hyperref}
\usepackage[normalem]{ulem}
\usepackage{dsfont}
\usepackage{float} 
\usepackage{fancyhdr}

\usepackage{amsmath}

\begin{document}

\title{Space-Time Quantum Metasurfaces} 

\author{Wilton J. M. Kort-Kamp}
\affiliation {Los Alamos National Laboratory, Los Alamos, NM 87545, USA}

\author{Abul K. Azad}
\affiliation {Los Alamos National Laboratory, Los Alamos, NM 87545, USA}

\author{Diego A. R. Dalvit$^*$}
\affiliation {Los Alamos National Laboratory, Los Alamos, NM 87545, USA}

\date{\today}

\begin{abstract}

Metasurfaces are a key photonic platform to manipulate classical light using sub-wavelength structures with designer optical response. Static metasurfaces have recently entered the realm of quantum photonics, showing their ability to tailor nonclassical states of light.  We introduce the concept of space-time quantum metasurfaces for dynamical control of quantum light. We provide illustrative examples of the impact of spatio-temporally modulated metasurfaces in quantum photonics, including the creation of frequency-spin-path hyperentanglement on a single photon and the realization of space-time asymmetry at the deepest level of the quantum vacuum. Photonic platforms based on the space-time quantum metasurface concept have the potential to enable novel functionalities, such as on-demand entanglement generation for quantum communications, nonreciprocal photon propagation for free-space quantum isolation, and reconfigurable quantum imaging and sensing.
\end{abstract}
  
\maketitle

\noindent

The generation, manipulation, and detection of nonclassical states of light is at the heart of quantum photonics. As quantum information can be encoded into the different degrees of freedom of a single photon, it is highly desirable to develop photonic platforms that allow to control them while maintaining quantum coherence. Metasurfaces \cite{Kildishev2013, Chen2016}
have recently transitioned from the classical to the quantum domain \cite{Solntsev2020} and enabled enhanced light-matter interactions facilitated by the ultrathin subwavelength nature of their constituent scatterers. Spin-orbital angular momentum entanglement of a single photon  
has been demonstrated \cite{Stav2018} using geometrically tailored metasurfaces that induce  spin-orbit coupling of light via the Pancharatnam-Berry phase \cite{Bomzon2002}, and multiphoton interferences and polarization-state quantum reconstruction has also been achieved in a single geometric phase metasurface \cite{Wang2018}. A metasurface-based interferometer has been demonstrated for generating and probing entangled photon states \cite{Georgi2019}, opening opportunities for implementing quantum sensing and metrology protocols using metasurface platforms. 
Embedded quantum building blocks into arrays of meta-atoms, including quantum dots, semiconductor emitters, and nitrogen-vacancy centers, result in enhanced Purcell factors
\cite{Vaskin2019}, directional lasing \cite{Xie2020}, and circularly-polarized single-photon emission \cite{Kan2020}.  Recently, quantum metasurfaces based on atomic arrays have been proposed \cite{Bekenstein2020}.

Most demonstrations of metasurfaces in quantum photonics are based on static meta-atoms whose optical properties are determined by their material composition and geometrical design that cannot be changed on demand.
A few realizations of active quantum metasurfaces have been reported, for example, for tuning spontaneous emission from a Mie-resonant dielectric metasurface using liquid crystals \cite{Bohn2018}.
However, a fully tailored response requires quantum metasurfaces that can continuously alter their scattering properties simultaneously in space and time. At the classical level, spatio-temporally modulated metasurfaces \cite{Shaltout2019}
have been shown to provide that higher degree of control, both by reconfigurable  
and fully-dynamic tailoring of the optical response of meta-atoms using
analog and digital modulation schemes \cite{Cardin2020,Zhang2018}. 

Here, we put forward the concept of space-time quantum metasurfaces (STQMs) for spatio-temporal control of quantum light. In order to highlight the broad implications of this  concept  in different areas of quantum science and technology, we discuss two instances of how STQMs operate both at the single-photon and virtual-photon levels. We describe STQM-enabled hyperentanglement \cite{Barreiro2005}  manipulation of nonclassical states of light and STQM-induced photon pair generation (Fig. 1) in a process analogue to the dynamical Casimir effect \cite{Moore1970}.

\vspace{0.5cm}
\noindent
{\bf Results:} 

We introduce the STQM concept by
considering the transmission of a single photon through a metasurface whose  meta-atoms are modulated in space and time. The metasurface is composed of identical anisotropic scatterers (Fig. 2a) suitably rotated with respect to each other. The combination of anisotropy and rotation results in circular cross-polarization conversion and a spin-dependent geometric phase distribution $\Psi({\bf r})$  akin to spin-orbit coupling. To minimize photon absorption the metasurface is assumed to be comprised of low-loss dielectric meta-atoms. The spatio-temporal modulation is modeled as
a perturbation of the electric permittivity, 
$\epsilon({\bf r}, t)=\epsilon_{um} + \Delta \epsilon \cos(\Omega t -\Phi({\bf r}))$, where $\epsilon_{um}$ is the unmodulated permittivity, $\Delta\epsilon$ the modulation amplitude, $\Omega$ the modulation frequency, and $\Phi({\bf r})$ is a ``synthetic" phase. 
Such type of modulation has been recently demonstrated using a heterodyne laser-induced dynamical grating on an amorphous Si metasurface via the nonlinear Kerr effect \cite{Guo2019},  setting a traveling-wave permittivity perturbation with $\Phi({\bf r})=\boldsymbol{\beta} \cdot {\bf r}$ ($\boldsymbol{\beta}$ is an in-plane momentum ``kick"). 
Note that the geometric phase is fixed by the design of the metasurface while the synthetic phase is reconfigurable on-demand. 

\begin{figure}[t!]
\includegraphics[width=1.0\linewidth]{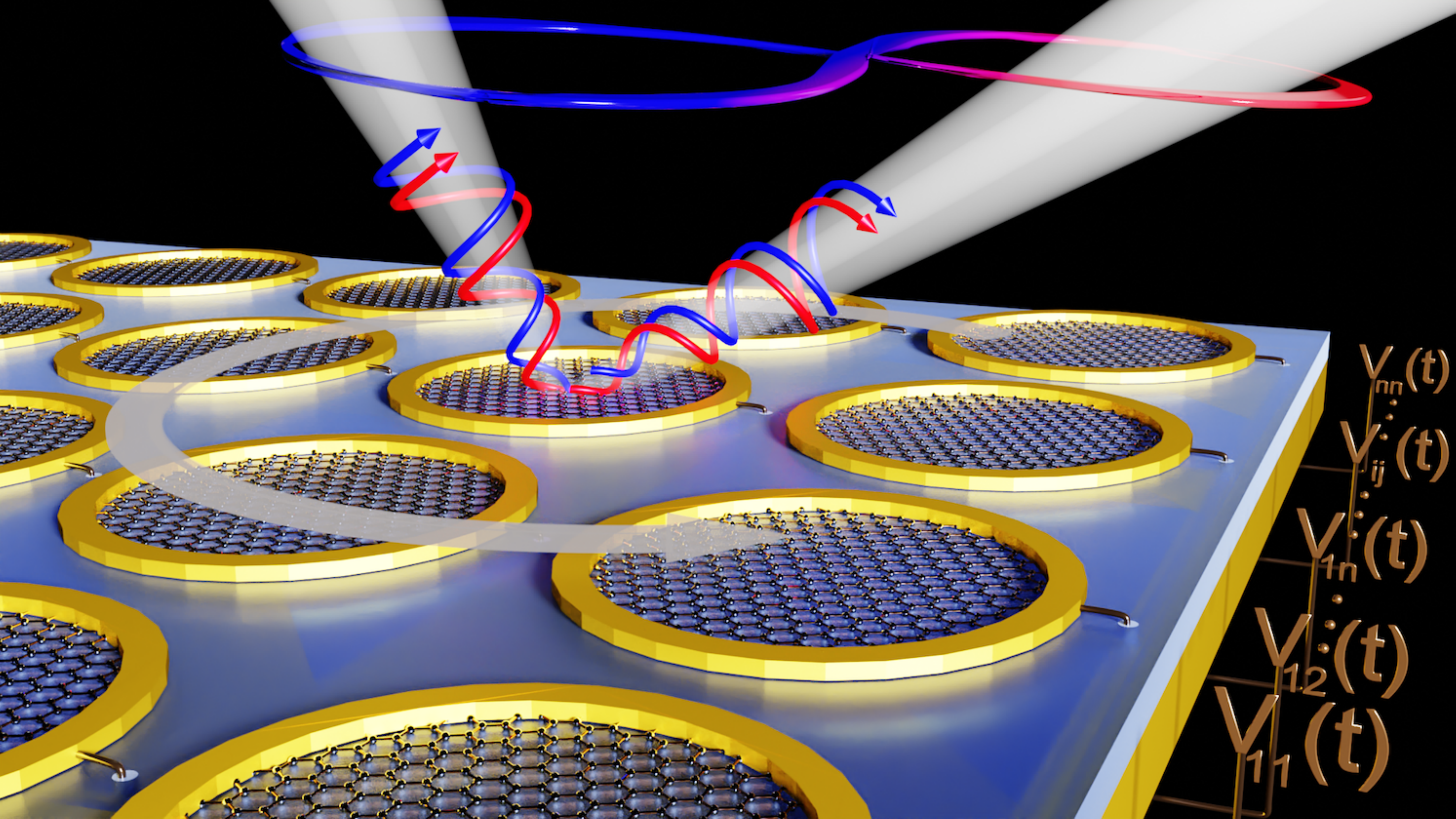}
\caption{{\bf Conceptual representation of a space-time quantum metasurface.}
A spatio-temporal spinning modulation of graphene nanostructures generates
entangled vortex photon pairs out of the quantum vacuum.
}
\label{Fig1}
\end{figure}

\vspace{0.5cm}
\noindent{\bf{STQMs for on-demand entanglement manipulation:}}
The geometry of the dielectric nanoresonator can be tailored so that 
it has maximal cross-polarized transmission (Fig. 2b) and at the same time so that 
its Mie electric and magnetic dipolar resonances dominate the optical response of the metasurface. One can then describe the interaction of each Mie resonator with light using the effective Hamiltonian $H_{int}=-{\bf p}\cdot{\bf E}-{\bf m}\cdot{\bf B}$ \cite{Kuznetsov2016,Novotny2007}, where ${\bf p}$ and ${\bf m}$ are the electric and magnetic dipole operators and ${\bf E}$ and ${\bf B}$ are the local quantized electromagnetic fields. Higher-order Mie resonances can be neglected because the transmissivity and reflectivity of the metasurface are well-described by that of an array of 
electric and magnetic dipoles corresponding to the two lowest Mie multipoles.
It is convenient to 
trace over the nanostructure's degrees of freedom to express the Hamiltonian only in terms of 
photonic modes by relating dipoles and fields via effective electric $ \boldsymbol{\alpha}_E$ and magnetic $ \boldsymbol{\alpha}_M$ polarizability tensors (see Supplementary Information for the derivation of the polarizabilities).
We show in Fig. 2c the relevant polarizability components for describing the coupling with the normally-incident photon and in Fig. 2d the electric field distributions at the Mie resonance frequencies. The unmodulated Hamiltonian describing cross-polarized transmission has an effective coupling strength $\alpha^{(cr)}_{um}(\omega)$ that is a simple combination of the 
electric and magnetic polarizabilities (see Methods). 

Upon spatio-temporal modulation,
the effective polarizabilities adiabatically follow the harmonic driving  because the response times of semiconductors ($<100$ fs for the nonlinear Kerr response time in amorphous Si) are much faster than THz modulations achievable with all-optical schemes.  Hence, 
\begin{equation}
\alpha^{(cr)}(\omega;{\bf r},t)=\alpha^{(cr)}_{um}(\omega) + \Delta\alpha^{(cr)}(\omega) \cos(\Omega t- \Phi({\bf r})).
\end{equation} 
We calculate the polarizability modulation amplitude  $\Delta\alpha^{(cr)}(\omega)$
from the dependency of transmissivity on permittivity modulation (Fig. 2e). For a $1\%$ permittivity modulation depth the resulting polarizability change is approximately $20 \%$, 
the  increase originating from the strong dispersion of the unmodulated polarizability close to the input frequency. 
The STQM Hamiltonian $H_{1}(t)$ is the sum of the unmodulated part plus a modulation contribution that 
annihilates the input  photon and creates a new one with  Doppler-shifted frequency and synthetic phase, in addition to flipping its spin components and adding geometric phases in the same way as the unmodulated part (see Methods). In this work we restrict to unitary evolution as photons do not suffer from severe decoherence problems and absorption is negligible in high-index dielectrics \cite{Stav2018}.

\begin{figure}[t!]
\includegraphics[width=1.0\linewidth]{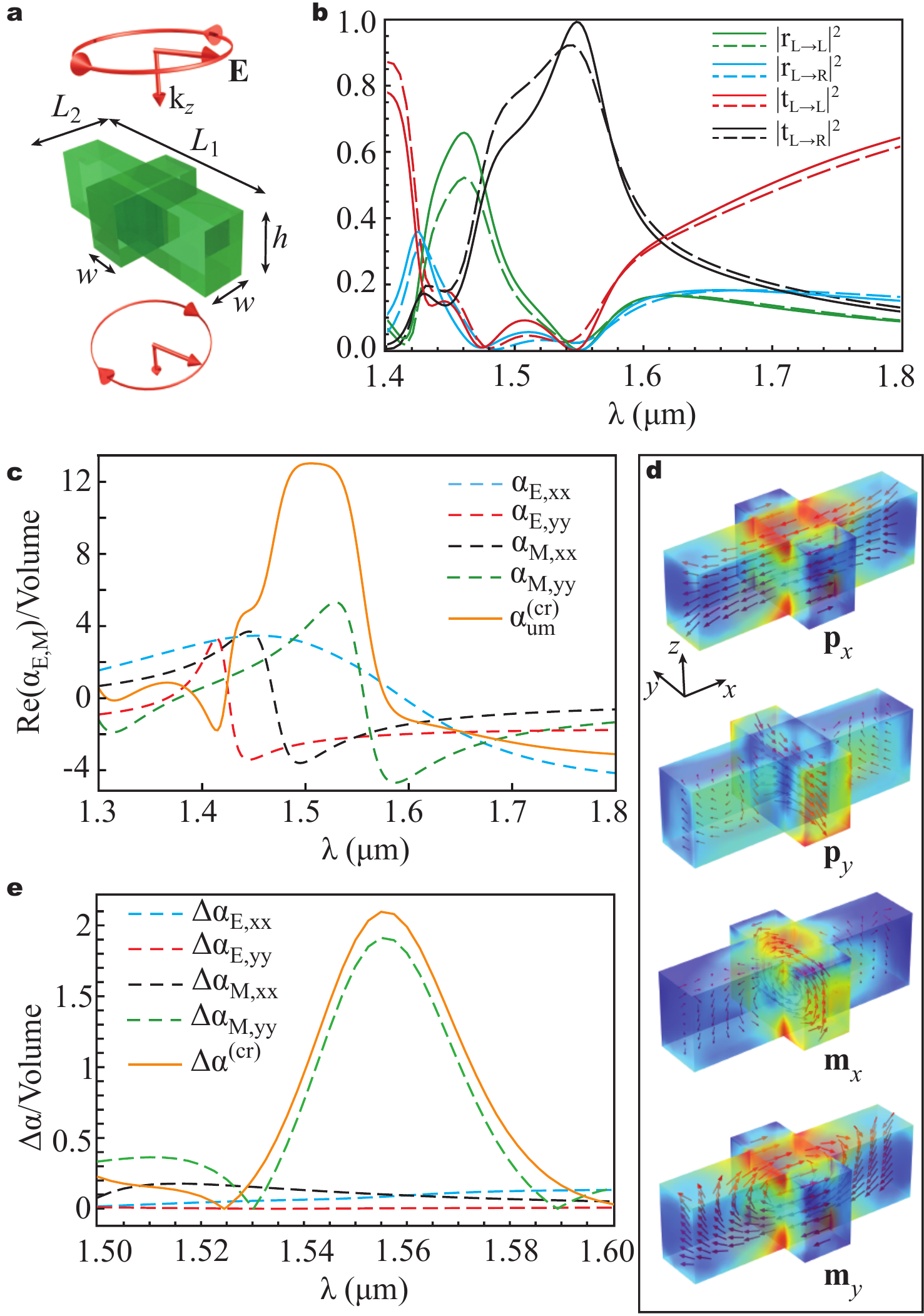}
\caption{{\bf Effective polarizabilities of all-dielectric space-time quantum metasurfaces.}
{\bf (a)} Anisotropic amorphous Si Mie nanocross meta-atom with optimized  
geometrical parameters for maximal cross-polarization transmission 
for a normally-incident $\lambda_{in}=1550$ nm input photon. Parameters are $L_1=950$ nm, 
$L_2=435$ nm, $h=300$ nm, $w=200$ nm, and square unit cell with period $P=1200$ nm.
{\bf (b)} Co- and cross-polarized reflectivity and transmissivity for the full metasurface (solid) and electric/magnetic dipole array (dashed).
{\bf (c)} 
Real parts of the electric and magnetic polarizabilities normalized by the meta-atom volume. 
Solid line is the effective unmodulated coupling strength for cross-polarized transmission:
$\alpha^{(cr)}_{um} \approx  0.6 \mu{\rm m}^3$ at the input frequency. 
{\bf (d)} Electric field distribution for the two electric and 
the two magnetic Mie resonances.
{\bf (e)} Polarizability  modulation amplitudes for permittivity modulation depth 
$\Delta\epsilon/\epsilon_{um}=1 \%$. Solid line 
is the polarizability modulation amplitude for cross-polarized transmission:  
$\Delta\alpha^{(cr)}/\alpha^{(cr)}_{um} \approx 0.2$ at the input frequency.
}
\label{Fig2}
\end{figure}

When the geometric phase is a linear function of the meta-atoms' positions it generates spin-momentum correlations, while a linear synthetic phase creates momentum-frequency correlations. The two correlations are intertwined through momentum and the photon evolves into a state that is hyperentangled in spin, path, and frequency
\begin{equation}
\!\!|\psi(t)\rangle \!=\! \sum_{p,q} [c^{(R)}_{p,q}(t) |\omega_{p}; {\bf k}_{p,q}; \!R\rangle + 
c^{(L)}_{p,q}(t) |\omega_{p}; {\bf k}_{p,-q}; \!L\rangle],
\label{eq1}
\end{equation} 
where $p$ are integers, $q=0,1$, $R(L)$ denotes right (left) circular polarization, 
$\omega_p=\omega_{in}+p \Omega$ are harmonics of the input frequency $\omega_{in}$, 
${\bf k}_{p,q}={\bf k}_{in}+ p \boldsymbol{\beta} +q \boldsymbol{\beta}_g$  are in-plane momentum harmonics of the in-plane input wave-vector ${\bf k}_{in}$, and 
$\boldsymbol{\beta}_g$ is the momentum kick induced by the linear geometric phase.
We will denote states in the first term as $(p,q,R)$ and in the second term as $(p,-q,L)$, highlighting that
the geometric-phase-induced momentum kicks for right- and left-polarized photons have opposite directions.
To calculate the probability amplitudes we consider a normally-incident single-photon pulse and assume modulation frequencies and in-plane momentum kicks much smaller than the input frequency and input wave-vector. 
Since the dielectric metasurface enables large polarizability modulation amplitudes for modest permittivity variations, it is possible for the input photon to transition to multiple frequency/momentum harmonics during its transit within the metasurface. For input linear polarization, the transition probabilities to states $(p,q,R)$ and $(p,-q,L)$ are identical and are given by
\begin{equation}
|c^{(R/L)}_{p,q}(t)|^2  = \frac{1}{2} \cos^2\Big(\frac{\omega_{in} t \alpha^{(cr)}_{um}}{2h P^2}\Big) 
J^2_{p}\Big(\frac{\omega_{in} t \Delta\alpha^{(cr)}}{2hP^2}\Big)
\label{probabilities}
\end{equation}
when $p$ and $q$ have the same parity; for opposite parity the cosine is replaced by a sine. $J_p(x)$ is the Bessel function and probabilities for $\pm p$ are the same (see Supplementary Information for details on the state evolution). 

\begin{figure*}[t!]
\includegraphics[width=1.0\linewidth]{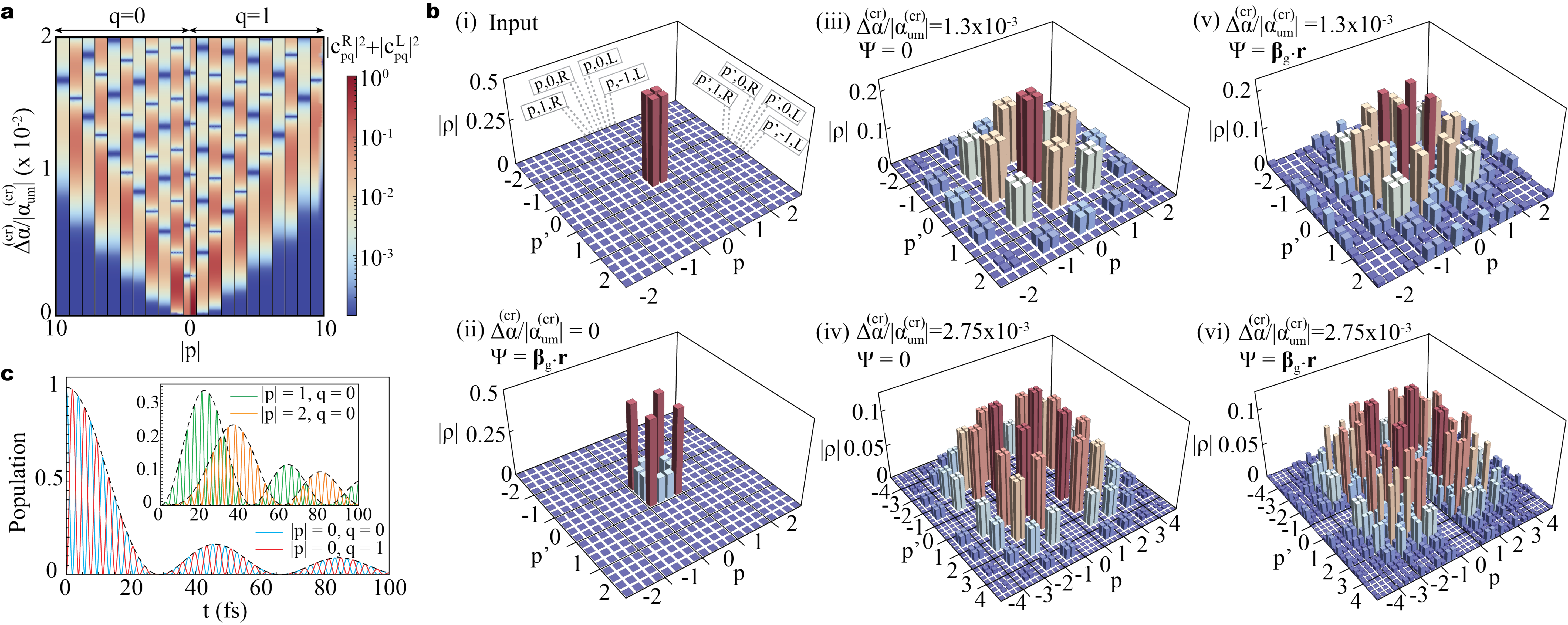}
\caption{
{\bf Entanglement manipulation with space-time quantum metasurfaces.}
{\bf (a)}  Conversion probability into an output photon in frequency harmonic $\omega_{in}+p\Omega$ and momentum harmonic $p \boldsymbol{\beta} +q \boldsymbol{\beta}_g$ versus polarizability modulation depth for the STQM of Fig. 2.
{\bf (b)} Density matrices of input (i) and output photon featuring
(ii) spin-path entanglement, 
(iii-iv)  frequency-path entanglement, and
(v-vi) frequency-spin-path 
hyperentanglement. Larger modulation depths result in more harmonics involved in the entangled states
(iv) and (vi).
{\bf (c)} Population dynamics of geometric-phase kicked $q=1$ and unkicked $q=0$ states for in-transit photon at $0.1$ modulation depth, showing Rabi oscillations for the fundamental frequency harmonic. Inset: Rabi dynamics in higher harmonics. Envelopes (dashed black) are the populations of
$p$-harmonics in the absence of geometric phase. 
}
\label{Fig3}
\end{figure*}

The probability that the output photon is in a given frequency/momentum harmonic as a function of the modulation depth is shown in Fig. 3a. At zero modulation, the output has the same frequency as the input and is approximately an equal superposition of right- and left-polarized geometric-phase-kicked states, with a small overlap with unkicked states due residual co-polarized transmission.
As the modulation increases, transitions to only the first few frequency/momentum harmonics occur and a larger amount of the available  Hilbert space is explored at large modulation depths. 
Figure 3b depicts the density matrices of the input (panel (i)) and  output photons for different configurations of the STQM, resulting in distinct kinds of quantum correlations: (ii) Geometric phase with spatio-temporal modulation off, giving a spin-path entangled output of same frequency as input;
(iii-iv) No geometric phase and spatio-temporal modulation on, resulting in frequency-path entangled cross-polarized output;
(v-vi) Geometric phase with spatio-temporal modulation on, delivering a frequency-spin-path hyperentangled output. 
It is possible to tailor the modulation depth to completely suppress the contribution of a given harmonic to the output state, as shown in (iv, vi) for the fundamental frequency. Under temporal modulation only, i.e., null synthetic phase (not shown), the output photon is unentangled (hyperentangled) in the absence (presence) of geometric phase.  Figure 3c shows the population dynamics  of different harmonics while the photon is in-transit inside the STQM. Interestingly, the evolution of populations with and without geometric phase are fundamentally different. Due to spin-orbit coupling the photon undergoes Rabi-flopping between state pairs
$(p,0,R)\leftrightarrow(p,-1,L)$ and $(p,0,L)\leftrightarrow(p,1,R)$, and this population exchange cannot take place at zero geometric phase. Unmodulated and modulated polarizabilities control the time-scales of  Rabi and synthetic-phase  dynamics, respectively.

When both phase distributions are azimuthally varying, i.e.,  
$\Psi({\bf r})=\ell_g \varphi$, $\Phi({\bf r})=\ell \varphi$ ($\ell_g$ and  $\ell$ integers), 
the  input photon becomes hyperentangled in frequency, spin, and orbital angular momentum (OAM) \cite{Calvo2006}. 
The state of the photon can be written as in 
Eq. (\ref{eq1}) replacing linear momentum harmonics ${\bf k}_{p,q}$ by  OAM harmonics $\ell_{p,q}=p \ell + q \ell_g$. 
Such a rotating synthetic phase could be implemented, e.g., via 
a heterodyne laser-induced dynamical grating with Laguerre-Gauss petal modes \cite{Eichler1986,Naidoo2012} to generate an all-optical spinning perturbation of the meta-atoms' refractive index.
As STQMs offer the possibility to reconfigure the synthetic phase on-demand, the question naturally arises as to what happens when the synthetic and geometric phase distributions have utterly different symmetry, for instance 
one is linear and the other spinning. It is then necessary to expand 
one phase in terms of a mode basis with symmetries appropriate for the other phase, e.g., plane waves  into cylindrical waves (see Supplementary Information for details of mixed-phase STQMs). 
As the synthetic phase creates frequency-path correlations and the geometric phase spin-OAM correlations, 
the two correlations are not intertwined and the STQM does not produce hyperentanglement but bipartite entanglement between pairs of degrees of freedom of the single photon. 
Finally, we mention that all the analysis presented in this section can be extended to other nonclassical inputs, such as two-photon Fock states.

\vspace{0.5cm}
\noindent{\bf{STQMs for tailored photon-generation out of quantum vacuum:}}
Space-time quantum metasurfaces can produce other nonclassical states of light and even induce nonreciprocity \cite{Sounas2017}
on quantum vacuum fluctuations. In addition to the photon-number-conserving Hamiltonian $H_1(t)$ discussed above, STQMs couple to the quantum electromagnetic field via a photon-number-non-conserving Hamiltonian $H_2(t)$  that creates photon pairs out of the quantum vacuum (see Methods). Their frequencies add up to the modulation frequency, $\omega+\omega'=\Omega$, thereby conserving energy, and this process is essentially an analogue of the dynamical Casimir effect (DCE) in which an oscillating boundary parametrically excites virtual into real photons 
\cite{Dalvit2006, Dodonov2010}. Although the mechanical DCE effect has not been detected  because it requires unfeasibly large mechanical oscillation frequencies,  various analogue DCE systems have been demonstrated \cite{Wilson2011, Jaskula2012, Lahteenmaki2013,Vezzoli2019}.
STQMs allow for a novel degree of dynamical control over the quantum vacuum through the synthetic phase:
The scattering matrix for the DCE process \cite{Maghrebi2013} becomes asymmetric via the spatio-temporal modulation, reflecting that Lorentz reciprocity is broken at the level of quantum vacuum fluctuations. 
\begin{figure*}[t!]
\includegraphics[width=1.0\linewidth]{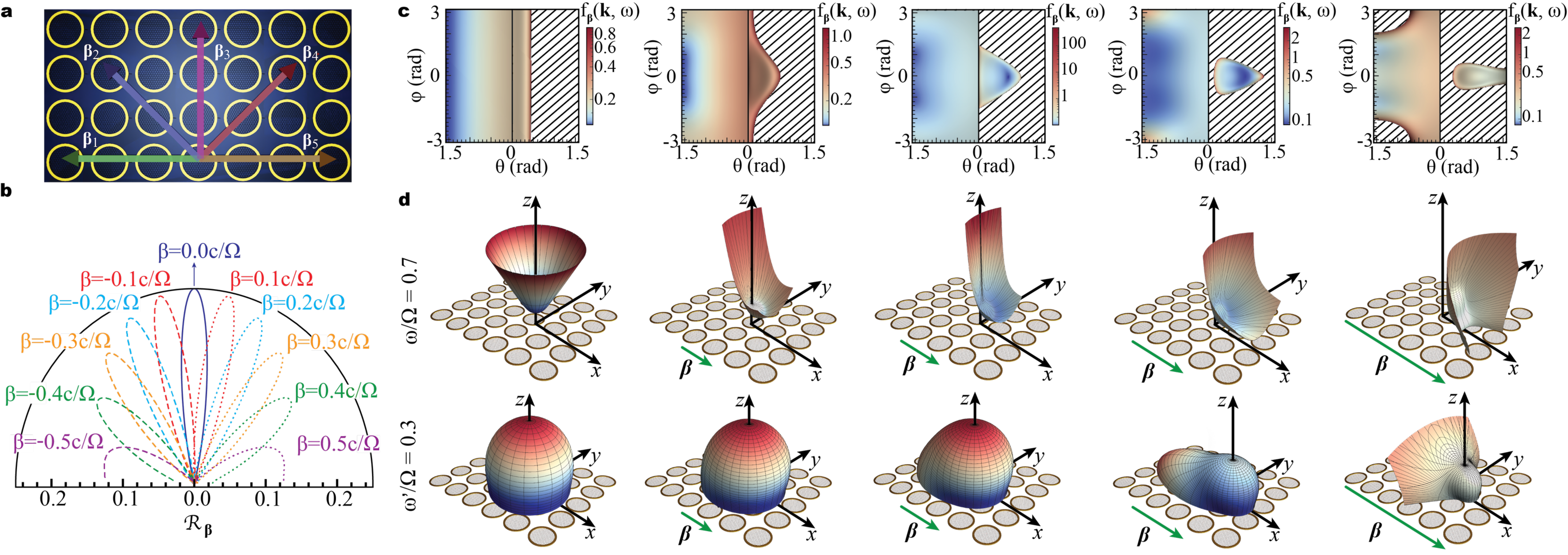}
\caption{{\bf Steered quantum vacuum.}
{\bf (a)} A linear synthetic phase is imprinted on a metasurface through a traveling-wave modulation and is tuned on demand (colored arrows) to steer the emitted dynamical Casimir photons.
{\bf (b)} Emission lobes of one photon for varying momentum kick and fixed (vertical) emission direction of its twin. 
{\bf (c)} Density polar plots of angular emission spectrum  
for various $\beta=(0, 0.2, 0.3, 0.38, 0.5) \Omega/c$
from left to right. The areas to the right (left) of the vertical solid line correspond to the angular emission spectrum of the high- (low-) frequency photon in a pair. Frequencies are $\omega/\Omega=0.7$ and 
$\omega'/\Omega=0.3$. Shaded zones correspond to forbidden photon emission directions.
Between the two rightmost panels two special events simultaneously happen:
the merge of the emission ``island" of the high-frequency photon with the grazing line and the birth of forbidden regions for the low-frequency photon.
{\bf (d)} Spherical polar plots for the same panels as in (c).
}
\label{Fig4}
\end{figure*}

\begin{figure}[t!]
\includegraphics[width=1.0\linewidth]{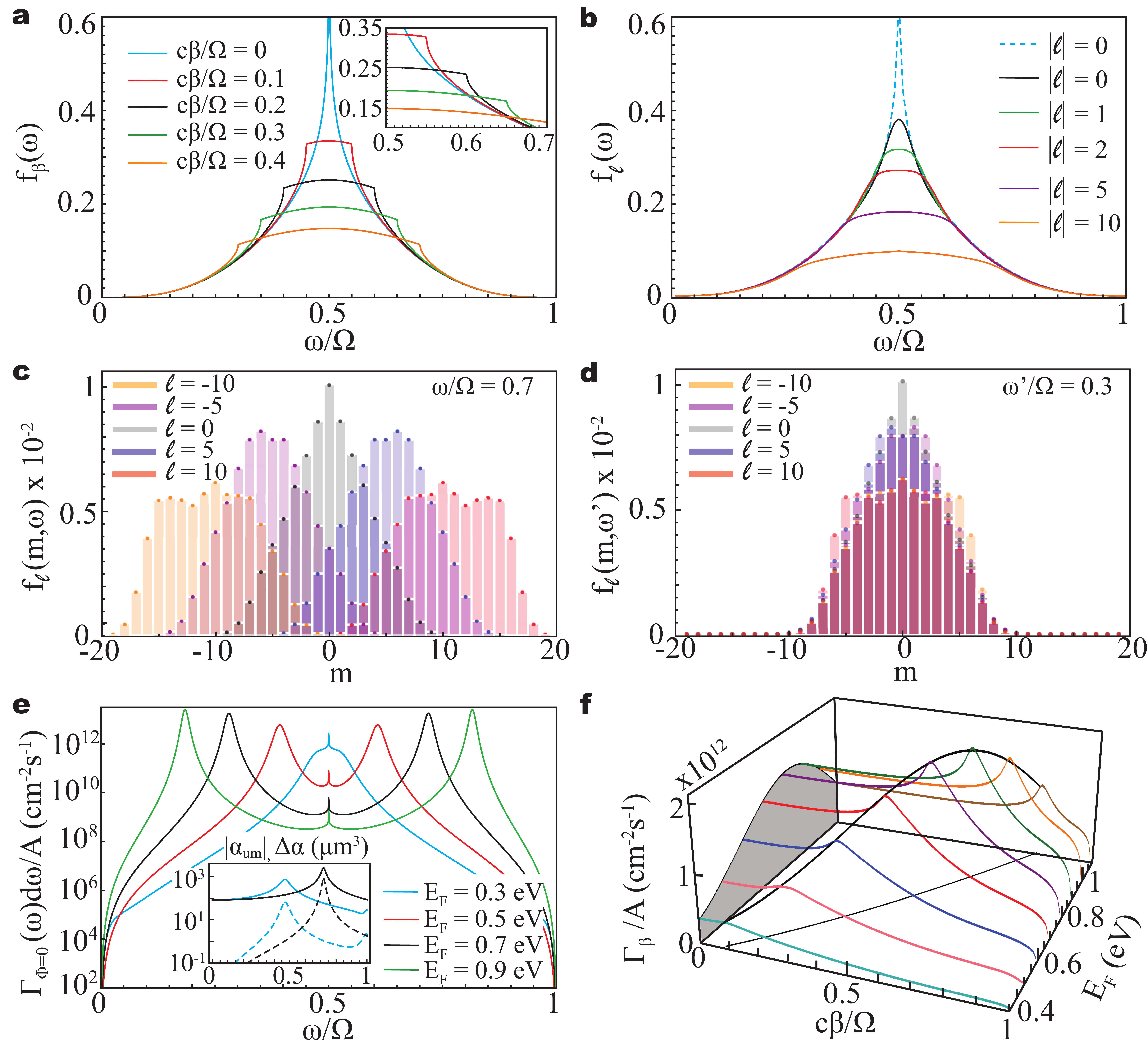}
\caption{{\bf Photo-emission rates for various synthetic phases.}
{\bf (a)} 
Spectral weight function for linear synthetic phase. 
Sharp edges of each plateau correspond to the special events of Fig. 4(c). Inset: Crossings 
responsible for non-monotonicity in (f).
{\bf (b)} 
Spectral weight function for rotating synthetic phase.
Solid lines correspond to a finite radius metasurface ($\Omega R/c=30$) and
dashed line is the $\ell=0$ case for an infinite metasurface.
{\bf (c)}  Angular-momentum spectra for finite radius metasurface for the high-frequency photon.
{\bf (d)} Same as (c) for the low-frequency photon.
{\bf (e)} 
Spectral photo-production rate for null synthetic phase 
for a graphene-disk STQM. Inset: unmodulated electric polarizability $\alpha_{um}(\omega)$ (solid) and modulation amplitude $\Delta\alpha(\omega)$ (dashed). 
{\bf (f)} 
Emission rate for linear synthetic phase. 
The profile on the left  shows the rate  at null synthetic phase.
The black thick curve joins peaks of maximal emission and the thin black curve 
$ c\beta = 2\omega_{res}(E_F)- \Omega$ is its projection on the $\beta-E_F$ plane.  
The rate decreases non-monotonically to zero at $\beta_{max}$.
In (e-f) parameters are: $\Omega/2\pi=10$ THz, $\Delta E_F/E_F=1 \%$, 
$n_{MS}=10^3 \; {\rm mm}^{-2}$, $D=5  \; \mu$m, and graphene mobility $\mu=10^4 \; {\rm cm}^2 \; {\rm V}^{-1} \; {\rm s}^{-1}$.
}
\label{Fig5}
\end{figure}

We first consider the case of the linear synthetic phase  (Fig. 4a) and set the geometric phase to zero. 
Momentum conservation dictates that the emitted photons must have in-plane momenta that add up to the imprinted kick, $\bf{k}+\bf{k}'= \boldsymbol{\beta}$, and the emitted photons are frequency-path entangled. In Fig. 4b we show  
the one-photon angular emission distribution for a fixed propagation direction of its twin, indicating how the externally imprinted momentum controls the directivity of the emission process. Figure 4c contains polar plots of the emission distributions for a given circularly-polarized photon pair (see Methods).
In the absence of kick, the high-frequency photon can be emitted in any azimuthal direction but it has a maximum polar angle of emission, while no such a constraint exists for the low-frequency photon.  As the magnitude  of the momentum kick $\beta$ increases, the distributions undergo intricate changes. 
The region of allowed emission for the first photon gets deformed when the kick is non-zero and at a critical value of the kick an ``island" of emission appears surrounded by a sea of forbidden emission directions (shaded areas). The island drifts to higher polar angles until it touches the grazing emission line, starts to shrink in size, and finally at  $\beta_{max}=\Omega/c$ it collapses to a point and the photon is only emitted parallel to the kick. Far-field emission above that value of the kick is not possible. Regarding the second photon, its emission distribution remains mostly unperturbed until two areas of forbidden emission appear at large polar angles and opposite to the kick direction. The forbidden region grows until it engulfs its allowed emission region and a second island forms (not shown). Finally, it ends up being emitted at a grazing angle but in a direction anti-parallel to the kick. The corresponding spherical plots are shown in Fig. 4d, with emission profiles resembling cone- (dome-) like shapes for the high- (low-) frequency photon and become increasingly distorted as the kick grows.  
When both photons are emitted with the same frequency, i.e., 
twin photons, the emission distribution is disk-shaped and gets elongated in a direction parallel to the kick as this increases in magnitude (not shown).
The modulation also excites  hybrid entangled pairs composed of one  photon and one evanescent surface wave  (shaded areas in Fig. 4c), and when $\beta>\beta_{max}$ only evanescent modes are created  and subsequently decay via non-radiative loss mechanisms. 

The  two-photon emission rate from an STQM of area $A$ with arbitrary  synthetic phase $\Phi({\bf r})$ is
\begin{equation}
{\Gamma}_{\Phi}= \frac{A n_{MS}^2 \Omega^4}{512 \pi^3 c^4} \! \int_0^{\Omega} \!\!\!d\omega  
|\Delta\alpha(\omega) + \Delta\alpha(\Omega-\omega)|^2 f_{\Phi}(\omega).
\label{rates}
\end{equation}
The rate scales as the square of the meta-atoms number surface density $n_{MS}$ indicating coherent emission of photon pairs. Electro-optical properties of the 
meta-atoms are contained in the modulated electric polarizability amplitude $\Delta\alpha(\omega)$. The  spectral weight function $f_{\Phi}$ results from the angular integration of all emission events
and is plotted in Fig. 5a for the case of the linear synthetic phase. 
$f_{\beta}(\omega)$ has a central plateau-like form with sharp edges that at zero kick coalesce into a single logarithmic integrable divergency at the center of the spectrum and corresponds to the emission of  twin photons \cite{MaiaNeto1996}. As the kick grows, the plateau becomes lower and at the maximum allowed kick the
spectral weight function vanishes.

STQMs can affect the quantum vacuum in more exotic ways, e.g., a modulation with a spinning synthetic phase ``stirs" the vacuum  (Fig. 1) and induces angular momentum nonreciprocity \cite{Sounas2013} at the level of quantum fluctuations. The rotating modulation generates vortex photon pairs that carry angular momenta 
satisfying $m+m'=\ell$. For $\ell\neq 0$ the average of the Poynting vector over all possible emission events results in a single vortex line along the synthetic spinning axis. 
Photon pairs are frequency-angular momentum entangled and their quantum correlations 
could be probed using photo-coincidence detection and techniques based on angular momentum sorting of light
\cite{Berkhout2010,Mirhosseini2013}.
The spectral weight function $f_{\ell}(\omega)$  is reported in Fig. 5b for a finite-radius metasurface, showing plateau-like structures as in Fig. 5a but without the sharp features on the edges, and with decreasing height as the spinning grows. There is a drastic but subtle difference between the two spectral weight functions $f_{\beta}(\omega)$ and $f_{\ell}(\omega)$ that is not apparent in the plots: The former vanishes beyond the finite kick threshold $\beta_{max}$, while there is no finite spinning threshold for the latter.
Figures 5c-d show the angular momentum spectra of  high- and low-frequency photons
in an emitted pair (see Methods). 
When the STQM does not imprint any spinning, the spectra are symmetric around the peak at $m=0$, with  oppositely twisted photons in each emitted pair.
When spinning is imprinted, the two spectra are related as $f_{\ell}(m,\omega)=f_{\ell}(\ell-m,\Omega-\omega)$ and the spectrum for the high- (low-) frequency photon is centered around $m=\ell$ ($m=0$). This is the angular-momentum equivalent of asymmetric linear momentum emission in Fig. 4d. 

Photo-emission rates can be boosted with suitable functional meta-atoms, such as
atomically-thin nanostructures made of plasmonic  materials that can support highly localized plasmons 
\cite{Yu2017,Abajo2015,Muniz2020} and enable large electric polarizabilities conducive to enhanced  coupling of the STQM with the quantum vacuum. As an example, we consider a STQM based on graphene disks whose Fermi energy $E_F$ is spatio-temporally modulated  (Fig. 1). Changing the Fermi energy it is possible to tune the plasmonic resonances into the DCE spectral range and to modify the  electric polarizability modulation amplitude (inset of Fig. 5e). Furthermore, the use of ultra-high mobility graphene samples minimizes photon absorption and
substantially enhances photo-production rates. 
Figure 5e depicts the spectral rate for a STQM for null synthetic phase
at selected Fermi energies, featuring Lorentzian peaks at complementary frequencies.
For high-Q resonances the emission rate for arbitrary synthetic phase can be approximated as
\begin{equation}
{\Gamma}_{\Phi}
\approx  g \Omega 
\; (A n_{MS}^2 D^6 \omega^4_{res} /c^4) \;  f_{\Phi;res} \Big(\frac{\Delta E_{F}}{E_{F}}\Big)^2
\Big(\frac{\Omega}{\gamma}\Big)^3 
\label{lorentz}
\end{equation}
Here, $g$ is a numerical factor determined by the plasmon eigenmode, 
$D$ the disk diameter, $\omega_{res}$ is the plasmonic resonance frequency, 
$f_{\Phi;res}$ the spectral weight on resonance,
$\Delta E_{F}/E_{F}$  the Fermi energy modulation depth,
and $\Omega/\gamma \gg 1$ with $\gamma$ the scattering rate of graphene.
(see Supplementary Information for the derivation of the polarizability and emission rate).
Figure 5f shows the emission rate for the linear synthetic phase as a function of momentum kick and
Fermi energy. Giant photon-pair production rates on the order of $10^{12}$ photons$/{\rm cm}^2 {\rm s}$
are obtained at low-THz driving frequencies and modest modulation depths. 
Conventional electrical doping may not allow to reach the large Fermi energies where the rate is maximized, 
but it suffices for exploring the lower Fermi energy region where photon-pairs are already produced
in giant numbers. 
A heterodyne dynamical grating based on ultrafast all-optical THz modulation of graphene conductivity \cite{Tasolamprou2019}
could enable giant and steered photo-pair emission out of the quantum vacuum. 
Finally, we note that electro-optical ultrafast on-chip graphene modulators \cite{Li2014,Phare2015,Kovacevic2018}  could potentially be employed to independently bias different STQM graphene pixels with designer temporal delays to implement complex synthetic phases.

\vspace{0.5cm}

\noindent
{\bf Discussion:} Metasurfaces are crossing the classical-quantum divide to offer novel possibilities for flat quantum optics and photonics. On the quantum side, they can become an enabler platform for generating and manipulating nonclassical states of light in real time.  We uncovered a key property of space-time quantum metasurfaces relevant for potential applications: On-demand reconfiguration of the synthetic phase allows dynamically tunable quantum correlations, enabling to tailor the nature of entanglement depending on the symmetry properties of both geometric and synthetic phases.  
We also illustrated a second key property of space-time quantum metasurfaces with fundamental relevance: 
Lorentz nonreciprocity at the deepest level of vacuum fluctuations is attained through joint space and time modulations of optical properties and can be interpreted as an asymmetric quantum vacuum.

Spatio-temporally modulated quantum metasurfaces have the potential to become a flexible photonic platform for generating nonclassical light with designer spatial and spectral shapes, for on-demand manipulation of entanglement for free-space communications, and for reconfigurable sensing and imaging systems. 
Conversion efficiencies into specific frequency, linear momentum, or orbital angular momentum harmonics for selective quantum information encoding could be enhanced through advanced modulation protocols. Incorporation of quantum matter building blocks into space-time metasurfaces may further expand the possibilities afforded by the proposed platform.    
As such, space-time quantum metasurfaces can provide breakthrough advances in quantum photonics.
\vspace{0.5cm}

\noindent
{\bf Methods:} 
The effective polarizability tensors $\boldsymbol{\alpha}_E(\omega)$ and $\boldsymbol{\alpha}_M(\omega)$ of the dielectric nanostructure are obtained using a Cartesian multipole expansion \cite{Evlyukhin2013} of the full-wave simulated electromagnetic field under a  plane wave excitation, and computing ratios of the resulting Mie electric and magnetic dipoles to the incident field at the nanostructure's center.
The Hamiltonian for the all-dielectric STQM in cross-polarized transmission is 
\begin{eqnarray}
&& H_1(t) = - \sum_{j,\gamma,\gamma'} \; [\alpha^{(cr)}_{um}(\omega) +\Delta\alpha^{(cr)}(\omega) \cos(\Omega t -\Phi_{\!j})] \cr
&& \times A^{*}_{\gamma;j} A_{\gamma';j} e^{i(\omega -  \omega')t} 
[e^{i\Psi_{\!j}}  a^{\dagger}_{\gamma,R} a_{\gamma',L} 
\! + \! e^{-i \Psi_{\!j}}   a^{\dagger}_{\gamma,L} a_{\gamma',R} 
] \! + \! h.c. \nonumber
\end{eqnarray}
The sums are over all meta-atoms and field modes, the geometric $\Psi({\bf r})$  and synthetic $\Phi({\bf r})$ phase distributions are evaluated at the position of the meta-atoms,
$\Omega$ is the modulation frequency, $A_{\gamma}$, $A_{\gamma'}$ are spatial modes, and $a_{\gamma',L/R}$ and $a^{\dagger}_{\gamma,R/L}$ are annihilation and creation operators of circularly polarized photons.
The unmodulated coupling strength is 
\begin{equation}
\alpha^{(cr)}_{um}(\omega)\!=\! {\rm Re}[\alpha_{E,xx}(\omega) + \alpha_{M,yy}(\omega)- \alpha_{E,yy}(\omega)- \alpha_{M,xx}(\omega)]
\nonumber
\end{equation}
and $\Delta\alpha^{(cr)}(\omega)$ is the modulation coupling strength obtained by replacing in the above equation  each effective polarizability  by its respective modulation amplitude.

The Hamiltonian for the all-plasmonic STQM for two-photon emission  is
\begin{eqnarray}
H_2(t) &=&  \frac{1}{8} \sum_{j,\gamma,\gamma'}  \sum_{\lambda,\lambda'} \;
[\Delta\alpha(\omega-\Omega) + \Delta\alpha(\omega'-\Omega)] \cr
&\times&
A^{*}_{\gamma;j} A^{*}_{\gamma';j} \;
e^{i\Phi_j} e^{i(\omega+\omega'-\Omega)t} \;
a^{\dagger}_{\gamma,\lambda} a^{\dagger}_{\gamma',\lambda'} + h.c.
\nonumber
\end{eqnarray} 
where we neglected multiscattering between meta-atoms \cite{Holloway2005}. $\lambda,\lambda'$ are polarization states of the two photons and $\Delta\alpha(\omega)$ is the modulated electric polarizability amplitude of the meta-atom computed with the plasmon wavefunction formalism
\cite{Yu2017}. For a graphene disk of diameter $D$ with a high-Q localized bright-mode plasmonic resonance 
\begin{equation}
\Delta\alpha(\omega) \!\approx\!
\frac{\pi^2 a_1^2 \alpha_{fs} c D^2 \Delta E_{F}}{512 \hbar \Omega^2} 
\frac{ (\gamma/2\Omega)^2 }
{[((\omega\!-\!\omega_{res})/\Omega)^2 \!+\!  (\gamma/2\Omega)^2]^2} .
\nonumber
\end{equation}
\\
$\omega_{res}(E_F)= (\alpha_{fs} c E_F/\pi |\xi_{1}| \hbar  D)^{1/2}$ is the resonance frequency of the lowest bright-mode, $E_F$ and $\Delta E_{F}$ are the Fermi energy and its modulation amplitude,
$\alpha_{fs}$ is the fine structure constant,   
$\gamma=e v^2_F/E_F \mu$  the scattering rate of graphene,  
$v_F$ the Fermi velocity,  and $\mu$ the mobility.
The eigenmode coefficients $a_1=6.1$ and $\xi_1=-0.072$ determine $g=5 \pi^4 a_1^4 \xi_1^2/2(512)^3$ in Eq. (\ref{lorentz}). Emission rates are computed with time-dependent perturbation theory.
Non-paraxial quantization of the electromagnetic field with angular momentum is employed for the STQM with rotating synthetic phase \cite{Enk1994}.
The spectral weight functions for the linear and spinning synthetic phases are respectively decomposed into in-plane linear momentum $f_{\boldsymbol{\beta}}({\bf k},\omega)$ and angular momentum ${f}_{\ell}(m,\omega)$ spectra
\begin{eqnarray}
f_{\boldsymbol{\beta}}(\omega) &=& \int d{\bf k} \;(c/\Omega)^2  \; 
(1-|c{\bf k}/\omega|^2)^{-1/2}  f_{\boldsymbol{\beta}}({\bf k},\omega),  \cr
f_{\ell}(\omega)&=& \sum_m {f}_{\ell}(m,\omega).
\nonumber
\end{eqnarray}
Explicit expressions for these spectra can be found in the Supplementary Information.


\vspace{0.5cm}
\noindent
{\bf Acknowledgements:} This work was supported by the DARPA QUEST  program. 
We are grateful to A. Efimov,  M. Julian, C. Lewis, M. Lucero, and A. Manjavacas for discussions. 

\vspace{0.5cm}
\noindent
{\bf Author Contributions:} D.A.R.D. and W. K.-K. conducted the theory work and A. K. A. analyzed experimental feasibility. All authors discussed the findings and contributed to writing the paper.

\vspace{0.5cm}
\noindent
{\bf Competing Interests:} The authors declare no competing interests.

\vspace{0.5cm}
\noindent{$^*$Correspondence: dalvit@lanl.gov}


\end{document}